# Microhexcavity Plasma Panel Detectors

A. Mulski, C. Ferretti, D. S. Levin, N. Ristow
*University of Michigan, Department of Physics*
450 Church Street, Ann Arbor, Michigan 48109 United States, USA

M. Raviv Moshe, Y. Benhammou, A. Das, E. Etzion, D. Reikher
*Tel Aviv University, Beverly and Raymond Sackler School of Physics and Astronomy*
Tel Aviv 69978, Israel

P. S. Friedman
*Integrated Sensors, LLC, Ottawa Hills, Ohio, 43606 United States*

Plasma panel detectors are a variant of micropattern detectors that are sensitive to ionizing radiation. They are motivated by the design and operation of plasma display panels. The detectors consist of arrays of electrically and optically isolated pixels defined by metallized cavities embedded in a dielectric substrate. These are hermetically sealed gaseous detectors that use exclusively non-hydrocarbon gas mixtures. The newest variant of these "closed-architecture" detectors is known as the Microhexcavity plasma panel detector (µHex) consisting of 2 mm wide, regular close-packed hexagonal pixels each with a circular thick-film anode. The fabrication, staging, and operation of these detectors is described. Initial tests with the µHex detectors operated in Geiger mode yield Volt-level signals in the presence of ionizing radiation. The spontaneous discharge rate in the absence of a source is roughly 3-4 orders of magnitude lower compared to the rates measured using low energy betas.

## 1. INTRODUCTION

Plasma Panel Sensors (PPS) are a type of micropattern gaseous radiation detector inspired by commercial plasma display panels (PDPs). PDPs are long-lifetime, low mass, thin, hermetically sealed structures which provide a mature industrial fabrication procedure for heavily pixelated devices. These attractive features of PDPs may be incorporated into PPS designs.

Prototype PPS detectors [1] were made from commercial DC monochromatic plasma display panels. These efforts were followed by in-house fabricated pixelated PPS detectors [2], where each pixel in a grid of closed-cell pixels is individually biased for a Geiger mode gaseous discharge initiated by a traversing charged particle. In this respect, each pixel operates as an independent detector.

The packing fraction of these first detectors was 18% with rectangular pixels of 1 mm x 2 mm size and a minimum pixel wall separation of 1.7 mm. The units were intentionally fabricated with large pixel separations to investigate pixel operation with minimal electronic cross talk and discharge spreading.

This report presents early results on a new generation of plasma panel closed-cell detectors referred to as microhexcavity-PPS (µHex). The µHex was designed with thinner cavity walls and an overall larger active area than the 1st generation microcavity detector. A 3D rendering of the pixel layout, gas channels, circular readout anodes with connecting metallized vias, and bus channels is shown in Figure 1.

## 2. DETECTOR DESCRIPTION AND OPERATION

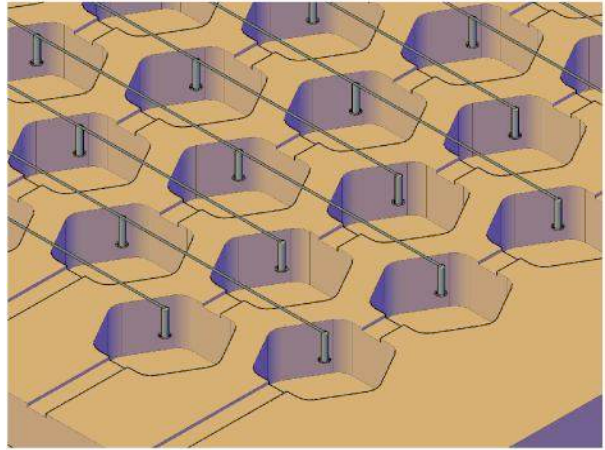

Figure 1: 3D µHex detector pixel AutoCAD design rendering with a transparent cover plate. Gas channels are shown as shallow grooves that bridge from pixel to pixel. Circular anodes with attached vias through the cover plate and connecting anode readout lines are also shown.

Pixels in the µHex detector (Figure 2) consist of 2 mm width metallized regular hexagonal cavities 1 mm deep in a 62.0 mm x 78.5 mm wafer of ceramic dielectric substrate. The metallized cavity serves as the cathode. High voltage is applied through a metal via which is connected to a HV bus line through surface mounted resistors shown in Figure 3. A thick-film printed anode is deposited on the bottom surface of the panel coverplate (glass or ceramic) and aligned to the center of the cavity. The readout (RO) anodes are connected to bus lines with metallized vias that run through the panel cover plate. These lines are orthogonal to the high voltage (HV) lines on the bottom side of the panel.

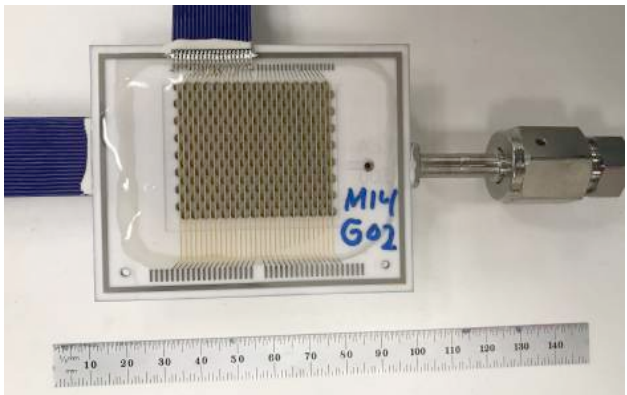 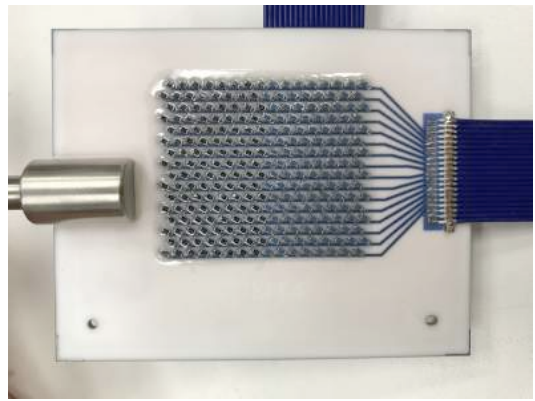

Figure 2 (left): A µHex with 32 RO lines with 8 pixels per line. 128 of these pixels are connected with parallel co-axial ribbon cables. The hexagonal structure of individual pixels is visible through the glass cover plate. The gas port is shown on the right of the detector. Figure 3 (right): 256 surface mount quench resistors (one for each pixel) on the back of the µHex.

Ion-pairs are created in the pixel gas volume when incident charged particles ionize the gas. This event initiates a Townsend avalanche which exceeds the Raether limit. Due to the high electric fields, this leads to a gas discharge limited in charge only by the pixel capacitance. Following discharge initiation, the E-field collapses and the discharge terminates. An example of the E-field structure from the first generation microcavity device is shown in Figure 4. The E-field gradient is largely determined by the anode geometry.

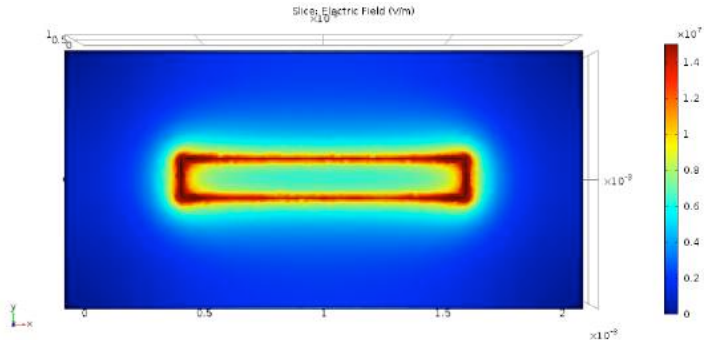

Figure 4: An E-field gradient at the top of a rectangular pixel in a 1st generation microcavity device created with COMSOL Multiphysics. The geometry of the anode is reflected in the structure of the E-Field. The strength of the E-field peaks around the edges and corners of the anode.

The time constant of the external quench resistor (*R* in Figure 5) and the pixel capacitance prevents the pixel from recharging before the gas ions neutralize. The pixel capacitance is less than 1 pF, and the quench resistance is on the order of hundreds of MΩs, yielding RC time constants of hundreds of microseconds. The signal is captured off the anode side of the panel from a low-value resistor between the RO line and ground. The signal is sent to a data acquisition system (TDC or scaler).

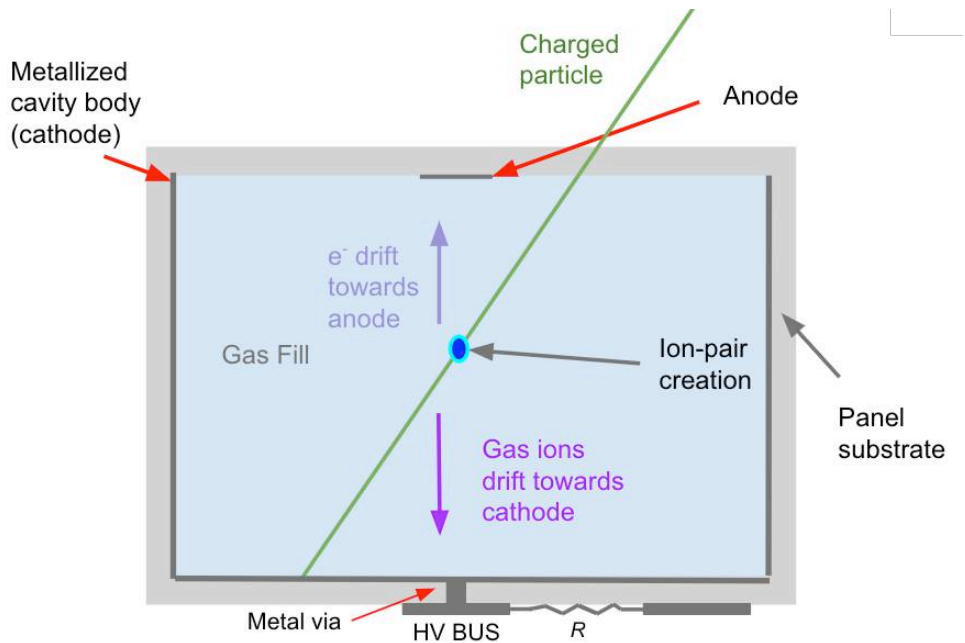

Figure 5: Pixel structure and representation of a charged particle entering the cavity.

## 3. RESULTS

A μHex detector filled to 740 Torr with a neon-based Penning gas mixture was used to characterize the signals and spatial response of these detectors. The voltage at which discharge pulses were first observed was approximately 900 V.

Beta particles from a $^{90}$Sr source were used to generate signals which were then characterized with a digital sampling oscilloscope (DSO). At these voltages, signals were not observed in the absence of the source. The DSO image in Figure 6 shows an envelope of 51 pulses. The pulse shape and amplitude varies little from pixel to pixel. The average pulse height is about 1 V and varies by ± 30% with a FWHM of 15 ns and a rise time of 7 ns.

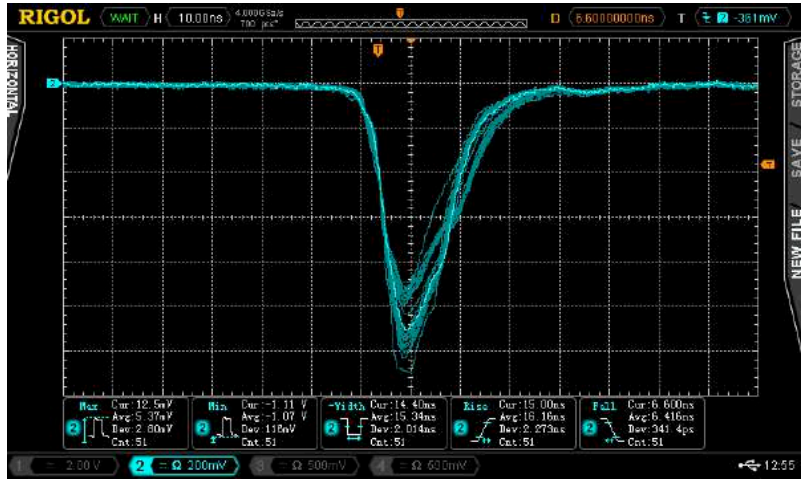

Figure 6: 51 signal pulses obtained from the μHex detector generated with βs from a $^{90}$Sr source. The signal shape is uniform across all pixels on the detector.

Hit rate dependence on high voltage was measured with the collimated $^{90}$Sr source. The source was positioned above the μHex such that each RO line received a different flux of β particles. The rate was measured at different values of applied high voltage beginning at the turn on voltage of 860 V and was increased in 10 V increments to 1000 V. The resulting voltage scan is shown in Figure 7. Each curve represents the hit rate measured on a RO line over equal time intervals as a function of voltage. The hit rates on all lines increase and begin to flatten as the voltage is increased. In the higher voltage range, residual rate increase is due to higher rates of afterpulsing. Each curve has a similar form, and the variation in overall rate is due to different exposures on each RO line. The spontaneous discharge rate with no source at 1000 V is less than 1 Hz per pixel.

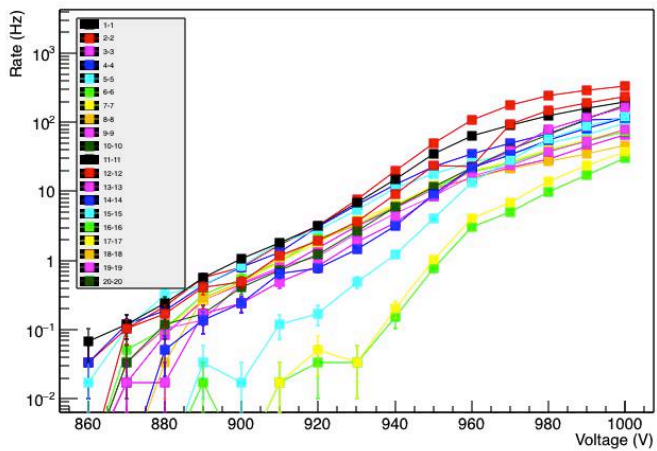

Figure 7: Rate as a function of applied HV with a $^{90}$Sr source. Each curve represents a single RO line. All RO lines received a different flux of β particles from the source. Although 20 RO channels are shown in the legend, only 16 of those channels are instrumented.

The spatial response and pixel isolation were investigated using a 1 mm diameter collimated [90]Sr source robotically scanned over the detector in fine steps. The source was placed 2 mm above the detector to reduce spray and was translated in X and Y in 250 micron increments. At each position, the hit rate on each RO line over 30s time intervals was measured. Signals from each readout channel were discriminated and the output was sent to a multichannel frequency counter. Figure 8 shows an example of a scan over a single RO channel as the collimated source is incremented across the row of pixels.

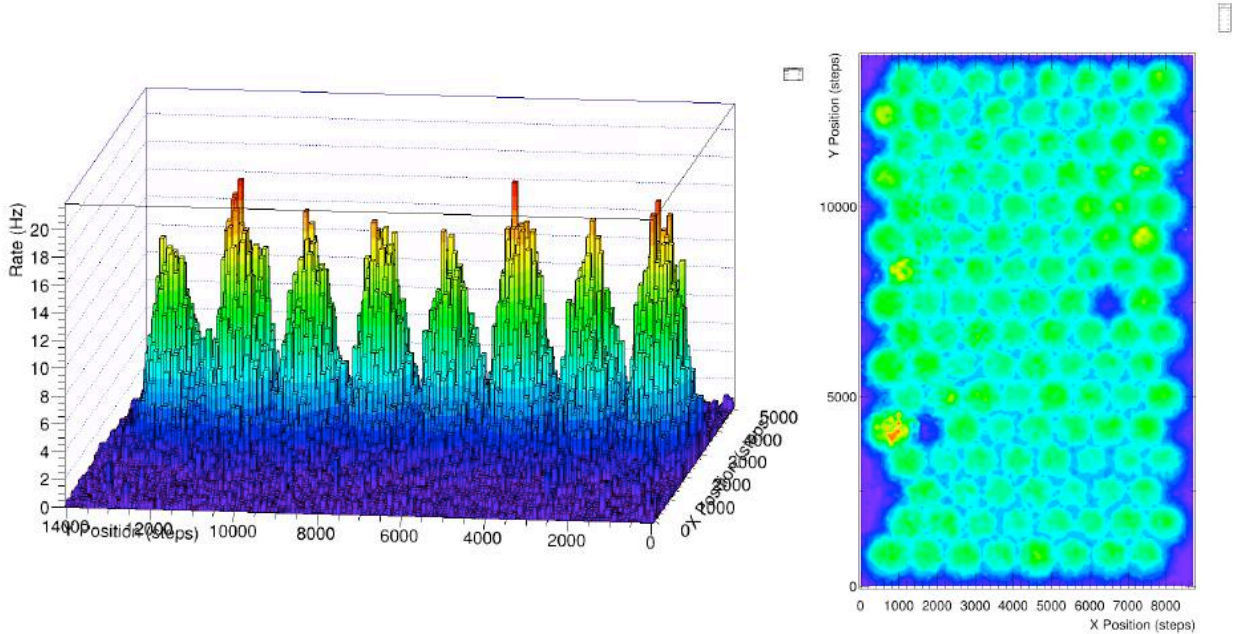

Figure 8 (left): Position scan over the μHex detector conducted with a collimated [90]Sr source. A portion of the scan over a single RO line is shown. Each step is 250 microns and the separation between pixel centers is 2.4 mm. The wall width between adjacent pixels is 0.4 mm. Figure 9 (right): Rate vs. position scan over the entire instrumented 1.9 cm x 3.3 cm μHex detector RO area with a collimated [90]Sr source. Each step is 250 microns.

The pixels on each RO channel, as shown in Figure 8, responded when irradiated by the source and were otherwise quiet, evident from the structure of the pixels visible in the plot. When the source is directly above the pixel, the hit rate is on the order of 20 Hz compared to much less than 1 Hz when the source is at the edges of the RO line. Since the collimator diameter was comparable to the radius of the pixel, the flux of incident β particles varied as the source was incremented across the RO line. The rate does not drop to zero between pixels as the source sprayed over the cavity walls into two neighboring pixels at once. The overall structure of the rate peaks indicates that each pixel operates as an isolated detection unit, uncoupled to neighboring pixels.

The position scan over the entire panel with 125 instrumented pixels (3 were disconnected) is shown in Figure 9. All instrumented pixels responded when irradiated by the source and did not respond otherwise. The hit rate was zero when the source was positioned over the edges of the detector, away from the pixels. The same β spillover effect observed between neighboring pixels in the position scan across a single RO line is visible in the scan over the entire panel. Figure 9 also shows three holes corresponding to unconnected pixels. These areas have the same hit rate as the background rate observed at the panel edges.

## 4. SUMMARY AND NEXT GENERATION DETECTOR

This report detailed the concept, fabrication, and first test results of the microcavity device referred to as the μHex detector. The pixels of these devices are individually biased for gaseous discharge and act as independent detectors, well isolated from one another. The first generation microcavity detector achieved a timing resolution of about 2.4 ns [2] and was operated for months on a single fill.

The μHex detector design achieved a higher packing fraction and spatial coverage than the 1st generation microcavity (i.e., 70% vs. 18% fill-factor), while maintaining no observable coupling between pixels. Future tests with the μHex detector will focus on measuring the efficiency and timing resolution of the device.

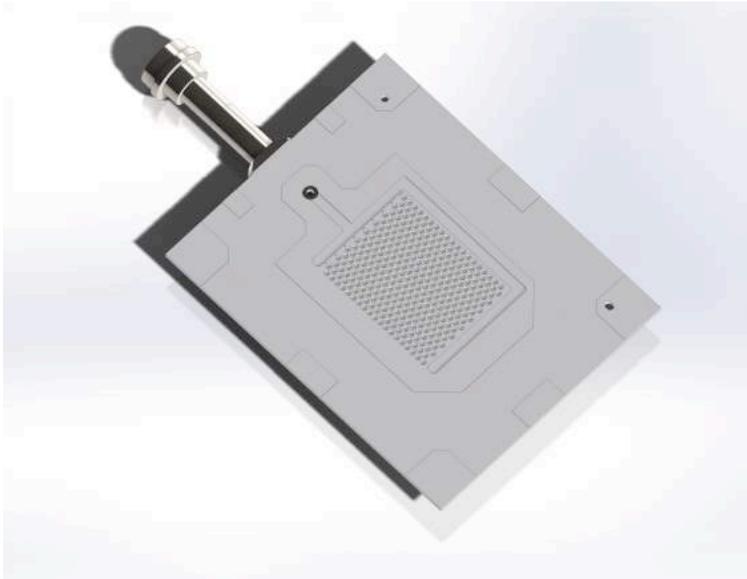

Figure 10: 3rd generation microcavity detector AutoCAD rendering. This detector will have circular pixels with 130 micron walls.

A higher fill factor of approximately 80% will be attempted for the 3rd generation microcavity detector. The design for the next generation detector includes implementing smaller circular cavities with 130 micron walls as shown in Figure 10. This design aims to achieve rates on the order of 100 KHz per $cm^2$ by packing close to 100 pixels per $cm^2$.

## Acknowledgments

This work was supported by the National Science Foundation under Award No. 1506117 and by the United States-Israel Binational Science Foundation under Grant No. 2014716. This work was also supported in part by the U.S. Department of Energy under Grant Numbers: DE-FG02-12ER41788 (Office of High Energy Physics) and DE-